\documentclass[useAMS,usenatbib]{mnras}
\usepackage{url,times,hyperref,graphicx,amsmath,amsfonts,amssymb,aas_macros,color,epsfig,epstopdf}
\usepackage[british]{babel} 

\title[Red nuggets in the present-day Universe]{Two new confirmed massive relic galaxies: red nuggets in the present-day Universe}

\author[A. Ferr\'e-Mateu et al.]{Anna Ferr\'e-Mateu$^{1,2}$,\thanks{E-mail: aferremateu@swin.edu.au (AFM)}
Ignacio Trujillo$^{3,4}$,
Ignacio Mart\'in-Navarro$^{5}$,
Alexandre 
\newauthor Vazdekis$^{3,4}$,
Mar Mezcua$^{6}$,
Marc Balcells$^{7,3,4}$
and Lilian Dom\'inguez$^{7,3,4}$ 
\\
$^{1}$Swinburne University of Technology, Hawthorn VIC 3122, Australia\\
$^{2}$Subaru Telescope, Hilo, HI 96720, USA\\
$^{3}$Instituto de Astrof\'{\i}sica de Canarias, La Laguna, E38205 Tenerife, Spain\\
$^{4}$Departamento de Astrof\'isica, Universidad de La Laguna, La Laguna, E38205 Tenerife, Spain\\
$^{5}$University of California Observatories, 1156 High Street, Santa Cruz, CA 95064, USA\\
$^{6}$D\'epartement de Physique, Universit\'e de Montr\'eal, C.P. 6128, Montreal, Quebec H3C 3J7, Canada\\
$^{7}$ Isaac Newton Group of Telescopes, Santa Cruz de La Palma, E38700 Tenerife, Spain \\
}

\date{Accepted for publication in MNRAS (January 18th 2017)}

\pubyear{2017}

\begin{document}
\label{firstpage}
\pagerange{\pageref{firstpage}--\pageref{lastpage}}
\maketitle

\begin{abstract}
We confirm two new local massive relic galaxies, i.e. untouched survivors of the early universe massive population: Mrk\,1216 and PGC\,032873. Both show early and peaked formation events within very short timescales ($<$1\,Gyr) and thus old mean mass-weighted ages ($\sim$13\,Gyr). Their star formation histories remain virtually unchanged out to several effective radii, even when considering the steeper IMF values inferred out to $\sim$3 effective radii. Their morphologies, kinematics and density profiles are like those found in the z$>$2 massive population, setting them apart of the typical z$\sim$0 massive early-type galaxies. We find that there seems to exist a \textit{degree of relic} that is related on how far into the path to become one of these typical z$\sim$0 massive galaxies the compact relic has undergone. This path is partly dictated by the environment the galaxy lives in. For galaxies in rich environments, such as the previously reported relic galaxy NGC\,1277, the most extreme properties (e.g. sizes, short formation timescales, larger super-massive black holes) are expected, while lower density environments will have galaxies with delayed and/or extended star formations, slightly larger sizes and not that extreme black hole masses. The confirmation of 3 relic galaxies up to a distance of 106\,Mpc implies a lower limit in the number density of these red nuggets in the local universe of 6$\times$10$^{-7}$ Mpc$^{3}$, which is within the theoretical expectations.
\end{abstract}

\begin{keywords}
galaxies: evolution -- galaxies: formation -- galaxies: kinematics and dynamics -- galaxies: stellar content -- galaxies: black holes
\end{keywords}



\section{Introduction}
The formation and evolution of massive early-type galaxies (ETGs) is a highly debated topic in modern astronomy. Two-phase formation models are becoming increasingly favored in the literature (e.g. \citealt{Naab2009}; \citealt{Oser2012}; \citealt{Hilz2013}; \citealt{RodriguezGomez2016}). First, the core of the galaxy is formed in-situ in a fast event at earlier epochs (z$\gtrsim$2), possibly starting with a top-heavy initial mass function (IMF) and evolving rapidly into a bottom-heavy one (e.g. \citealt{Vazdekis1997}; \citealt{Weidner2013}; \citealt{Ferreras2015}). This is followed by an accretion phase where the galaxy undergoes a series of random encounters with surrounding lower mass satellites. While the first phase is a dissipative event with extremely high star formation rates, the second phase is mostly driven by dry mergers, which place most of the newly accreted material at the periphery of the central massive core. This will add some stellar mass, changing the galaxy shape and making it larger, but leaving the center virtually untouched. This mechanism successfully explains the strong size and morphological evolution that is seen for the average population of massive galaxies since z$\sim$2, while keeping the stellar masses and velocity dispersions almost unvaried  (e.g. \citealt{Daddi2005}; \citealt{Trujillo2007}; \citealt{Buitrago2008}; \citealt{Cenarro2009}; \citealt{vanderWel2011}; \citealt{Toft2012}; \citealt{Belli2014}). However, some extra observational constraints have to be met if this picture is correct. For example, due to the stochastic nature of mergers, it is predicted that a few of these primordial massive galaxies should avoid all interactions, remaining untouched over cosmic time \citep{Quilis2013}. \\
The hunt for these descendants, nick-named \textit{massive relic galaxies}, has not been an easy endeavor. They have been found in the local universe with a variety of number densities (e.g. \citealt{Trujillo2009};  \citealt{Taylor2010}; \citealt{Valentinuzzi2010}; \citealt{Poggianti2013}; \citealt{Saulder2015}; \citealt{Tortora2016}). The main problem when comparing these studies resided in the different criteria used to define a relic galaxy: i.e. where the limits of compact and massive are set. While some were devoted to find the most extreme cases, both in size and in terms of no posterior accretion (e.g. \citealt{Trujillo2009}; \citealt{FerreMateu2012}), others would allow for some mass evolution and a more relaxed size constraint (e.g. \citealt{Valentinuzzi2010}; \citealt{Poggianti2013}; \citealt{Damjanov2014}). In fact, a fully detailed study of the structural and stellar population properties for a sample of such galaxies from \citet{Trujillo2009} rejected those first candidates, as they showed non-negligible episodes of recent star formation (e.g. \citealt{FerreMateu2012}; \citealt{Trujillo2012}; see also \citet{Damjanov2014} at intermediate redshifts). Nonetheless, the first confirmed relic was finally reported by \citet{Trujillo2014} (T14 hereafter), after a detailed analysis of its morphology, structural profiles, kinematics and stellar populations out to several galactocentric distances. That galaxy was NGC\,1277 and it had caught the attention of the extragalactic community for hosting an unprecedentedly large super massive black hole (SMBH; \citealt{vandenBosch2012}, but see \citealt{Emsellem2013}). \\
The claim of NGC\,1277 as the first fully confirmed massive relic galaxy led to two other interesting findings. On one hand, a detailed analysis of NGC\,1277 IMF proved that this galaxy demands a very steep IMF slope \citep{MartinNavarro2015b}. This steep IMF slope is in agreement with the recent claim for a tight relation between IMF slope and galaxy velocity dispersion (e.g. \citealt{Cappellari2012}; \citealt{LaBarbera2013}; \citealt{Ferreras2013}; \citealt{Spiniello2014}). Intriguingly, this galaxy did not show the predicted radial variations already seen for massive galaxies \citep{MartinNavarro2015a}. This is, although this galaxy did show a strong velocity dispersion gradient, its IMF remained virtually invariant with radius, making unclear what property really drives IMF variations \citep{MartinNavarro2015c}.\\
On the other hand, a dozen new massive relic candidates with similar properties to NGC\,1277 were reported, all seemingly hosting such extreme SMBHs \citep{vandenBosch2012} and thus being extreme outliers in the SMBH-galaxy local scaling relations. In order to understand if there was a connection between being a relic galaxy and hosting a \"ubermassive SMBH, we proposed an scenario to explain such deviations (\citealt{FerreMateu2015}; AFM15). In it, the host galaxy and the SMBH evolve rapidly up to z$\sim$2. After that point, the SMBH is fully assembled (e.g. \citealt{Barber2016}; \citealt{DiMatteo2016}) and one would expect the individual galaxy to grow partially in stellar mass during the accretion phase. However, as for the definition of massive relics, which implies no later accretion, these particular objects skip the second phase. This leaves them as natural outliers in the SMBH mass-galaxy host mass scaling relation. In order to prove it, we studied in AFM15 how much a sample of SDSS relic candidates would grow in stellar mass and velocity dispersion if such accretion phase were to be allowed and how this would move them along the SMBH-galaxy mass planes. Our results showed that this would, indeed, place them closer to the local scaling relations, as if they had followed the normal evolutionary path of massive galaxies. \\

At this point, the issue that needs to be tackled is whether NGC\,1277 is a rare and unique galaxy or, instead, a simple anecdotic example of this galaxy family, the massive relics. It is thus time to study in full detail the peculiar properties of other candidates, in order to further asses the nature of such interesting objects. It is clear by now the important role that the second phase of formation has on shaping galaxy evolution as we understand it today. However, as this phase is mostly accountable in the outer regions of the galaxies, studies of those properties need to be proven out to several galactocentric distances. \\
Some recent works have enlarged the samples of nearby compact massive candidates (e.g. \citealt{Saulder2015}; \citealt{Tortora2016}) and surveys such as the \textsc{hetmgs} \citep{vandenBosch2015}, although not aimed at their hunt, have also proven to be a good place where to find such massive relic candidates. In this work we will focus on two of the candidates in the  \textsc{hetmgs} survey, analyzing in full detail the structural, kinematical and stellar populations properties in a radial basis of PGC\,032873 and Mrk\,1216.  Besides confirming them as massive relics, the analysis performed here also allows us to deepen our understanding on the nature of their SMBHs. Most importantly, they both reside in different environments than the rich Perseus cluster of NGC\,1277, which will also allow us to discuss the possible impact of local environment. \\
Throughout this work we adopt a standard cosmological model with the following parameters: $H_0$=70 km s$^{-1}$ Mpc$^{-1}$, $\Omega_m$=0.3 and $\Omega_\Lambda$=0.7.

\section{Data and reduction}
We look for candidates to be massive relics in the \textsc{hetmgs} survey of \citep{vandenBosch2015}. This is a compilation of about a thousand nearby galaxies with resolved kinematics and SMBH mass estimates and it also includes galaxy sizes from the 2MASS Extended Source Catalog \citep{Jarrett2000}. This compilation is suitable for our work as it is biased towards dense galaxies and large velocity dispersions, such as NGC\,1277, but with no morphological constraints. According to their sample, 48 galaxies are compact  (R$_{\mathrm{e}} <$ 2\,kpc) and their stellar masses are above 10$^{11}$ M$_\odot$, which is the broad definition for a massive compact galaxy, i.e. red nugget, in the local universe (T09). We were awarded with time to observe two candidates: Mrk1216 and PGC032873. From the sample of 48 candidates, we focused on those that were visible during the period and that were not in a cluster (for the environmental study). While NGC\,1277 is a well studied galaxy, our two final candidates were both scarcely studied and not much information was available for them. Mrk\,1216 is an isolated galaxy, with only two closer galaxies at 1\,Mpc distance, that has been analyzed by \citet{Yildirim2015}. The authors showed that it is a compact, fast rotating, oblate early-type galaxy with no signs of substructure. Its peaked velocity dispersion is indicative of a high concentration of mass in its center, as proven by the large black hole mass they derived with dynamical models. In addition, the authors report that this galaxy should be embedded in an extended stellar halo, which would discard the hypothesis of being a stripped galaxy. Finally, a very recent work by \citet{Walsh2016} has presented a more detailed dynamical modeling for the SMBH of this galaxy, giving a more accurate estimate of its black hole mass. PGC\,032873 is a field galaxy firstly reported in \citet{vandenBosch2012} and later employed in AFM15 for the SMBH-massive relic study using SDSS data. However, there is no detailed dynamical study for it yet, having only an upper limit for its SMBH \citep{vandenBosch2015}. We use the sizes and stellar masses that have been derived in the above stated works (see Table 1). Throughout this paper we will use NGC\,1277 and its already published properties (e.g. \citealt{Trujillo2014}, \citealt{MartinNavarro2015b}) as our role model for massive relic galaxy, comparing our two new candidates with it. We will specify, nevertheless, if new values have been measured in order to have a consistent analysis.

\subsection{HST Imaging}
High resolution imaging for PGC\,032873 and Mrk\,1216 were obtained using the WFC  Advanced  Camera  for  Surveys  filter F160W  from the NASA/ ESA Hubble Space Telescope (HST) archive. The NGC\,1277 image was obtained with the F625W  (Sloan r) filter. These observations are associated with the programs GO:10546 (PI: Fabian) and GO:13050 (PI: van den Bosch). Figure 1 shows the stamps of the relic candidates to highlight their similarities and show their immediate neighborhood. 
\begin{figure*}
\includegraphics[scale=1.1]{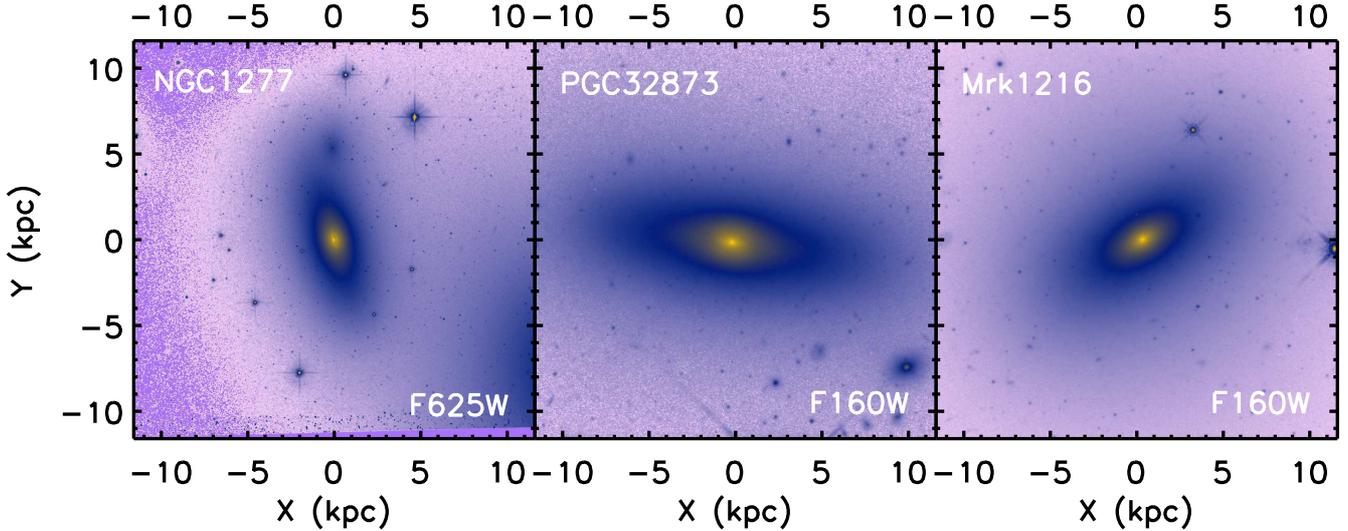}\\
\label{figure:1}
\caption
{The neighborhood of the relic candidates as seen by HST. The pixel scale of the F625W and F160W images are 0.05 arcsec and 0.06 arcsec, and the point-spread  function  FWHM  are  0.1 and 0.15 arcsec, respectively. They both resemble morphologically to NGC\,1277 (left panel), all showing highly symmetric shapes with no distortions neither bright tidal streams surrounding them. Both PGC\,032873 and Mrk\,1216 are field galaxies, with their closest neighbor further than 1Mpc distance.}
\end{figure*}

\subsection{Spectroscopy}
Two different spectroscopic observations were performed, one covering the blue spectral range to derive the stellar populations and kinematics with the ISIS spectrograph on the 4.2m William Herschel Telescope and the other covering the red spectral range to study the IMF variations with the FOCAS spectrograph on the 8m Subaru Telescope.\\
In the blue range, the blue grating R300B was centered at 5300\AA   with a $\times$2 binning in the spatial and spectral directions and the 1.5\arcsec\ slit placed along the major axis of each galaxy. This provided a spectral resolution of 5.2\AA. Several exposures of 30 minutes were taken for each galaxy, giving a total time on source of 2.5h (PGC\,032873) and 3.5h (Mrk\,1216), as stated in Table\,1. For the red side, the high resolution grism VPH850 was used with a $\times$2 binning in the spatial and spectral directions and the 0.6\arcsec\ slit was placed along the major axis of each galaxy. In this case the spectral resolution was 2.34\,$\rm\AA{}$, with total exposure times of 3 hours for each galaxy. In addition, spectrophotometric stars were observed with both configurations to perform a relative flux calibration.\\ 
Data reduction was performed with {\tt REDUCEME} \citep{Cardiel1999}, a reduction package optimized for long-slit spectroscopy that gives the errors that are propagated in parallel to the reduction process. We performed a typical data reduction: bias subtraction and flat-field correction, cosmic ray removal, sky subtraction, wavelength calibration, S-distortion, telluric lines, extinction corrections and relative flux calibration. We first corrected from both the systemic and the rotation velocity. Then, the spectra were summed up in annular radial bins to reach the required S/N for each study. For the stellar populations, a minimum S/N$\sim$20 is required, allowing to extend our analysis up to $\sim$\,4\,R$_{e}$ (effective radius). The IMF analysis requires higher S/N, at least of $\sim$100, which allowed us to extend to almost 3\,R$_{e}$ for the new two galaxies, almost twice than accomplished for NGC\,1277 in \citet{MartinNavarro2015b}.

\begin{table*}
\label{table:1}                      
\centering
\caption{Structural properties and published information}    
\begin{tabular}{c | c c c c c c c c}   
\hline\hline      
Galaxy         & Obs. date   &T$_{\mathrm{exp}}$& redshift  & R$_{\mathrm{e}}$ & R/R$_{\mathrm{e}}$& R$_{\mathrm{e}}$/R$_{\mathrm{shen}}$& $ \sigma$ at R$_{\mathrm{e}}$ &  M$_{*}$  \\
                    & (blue/red)    &    (hours)                &               &   (kpc)                   &              &                    &($\mathrm{km\,s^{-1}}$)  &  (10$^{11}$ M$_\odot$)\\   
\hline  
NGC\,1277     & 2013/2014 & 3.0/3.0 & 0.0169 & 1.2$\pm$0.1  &  3.5 & 0.26 & 385$\pm$6 & 1.3$\pm$0.3\\
PGC\,032873 & 2014/2015 & 2.5/3.0 & 0.0249 & 1.8$\pm$0.2  &  3.7 &  0.23 & 358$\pm$5 & 2.3$\pm$0.9\\
Mrk\,1216       & 2015/2015 & 3.5/3.0 & 0.0213 & 2.3$\pm$0.1  &  4.0 &  0.33 & 368$\pm$3 & 2.0$\pm$0.8\\

\hline                                
\end{tabular}

{Summary of the physical properties and other relevant information for the two relic candidates and  NGC\,1277. Columns 2 and 3 show the spectroscopic observational program and the total exposure times. Column 4 shows their redshift, while columns 5, 6 and 7 show the effective radii measured as described in Section 3.2, the radial coverage achieved for each galaxy and the departure from the local mass-size relation by \citet{Shen2003}. Columns 8 and 9 are the velocity dispersion within 1R$_{\mathrm{e}}$ apertures we measured with {\tt pPXF}. Column 10 shows the stellar mass published previously: \citet{Trujillo2014}, \citet{FerreMateu2015}, and \citet{Yildirim2015}, respectively.}
\end{table*}

\section{ANALYSIS: Confirmation as massive relic galaxies}
The most basic definition of a massive relic candidate stands for a galaxy that has not suffered any type of alterations since its early formation. Under the assumed two-phase formation scenario depicted here, we assume that the galaxy remains untouched after the first in-situ phase has finished, skipping the accretion phase \citep{Trujillo2009}. This implies that most of its observable properties must remain almost unvaried, showing a close resemblance to the high-z red nugget population. Therefore, in the case of the massive galaxy population, we will demand them to be massive, compact and with old stellar populations. Under the two-phase formation scenario, there should exist a transition zone where the properties of the population change from in-situ to accreted. Consequently, we must ensure that the massive relic criteria are sustained throughout the entire galaxy structure.

\subsection{Kinematics}
We measure the radial velocity and velocity dispersion before we bin the spectra along the slit to reach the desired S/N ratio using the software {\tt pPXF} \citep{Cappellari2004} with the MILES Library of stellar spectra \citep{SanchezBlazquez2006}. The spectra have been checked for emission filling with the code {\tt GANDALF} \citep{Sarzi2006}. \\
None of the two candidates show emission features in the relevant absorption lines (e.g. Balmer lines), reinforcing the hypothesis of old, passive evolving systems. Figure 2 shows the extracted radial kinematics of our relic candidates. Both show a strong radial velocity profile and a high centrally-peaked velocity dispersion (Table 1), dropping strongly within the effective radius. These kinematical features closely follow the behavior of NGC\,1277 and of the massive high-z counterparts (T14).

\begin{figure}
\centering
\includegraphics[scale=0.60]{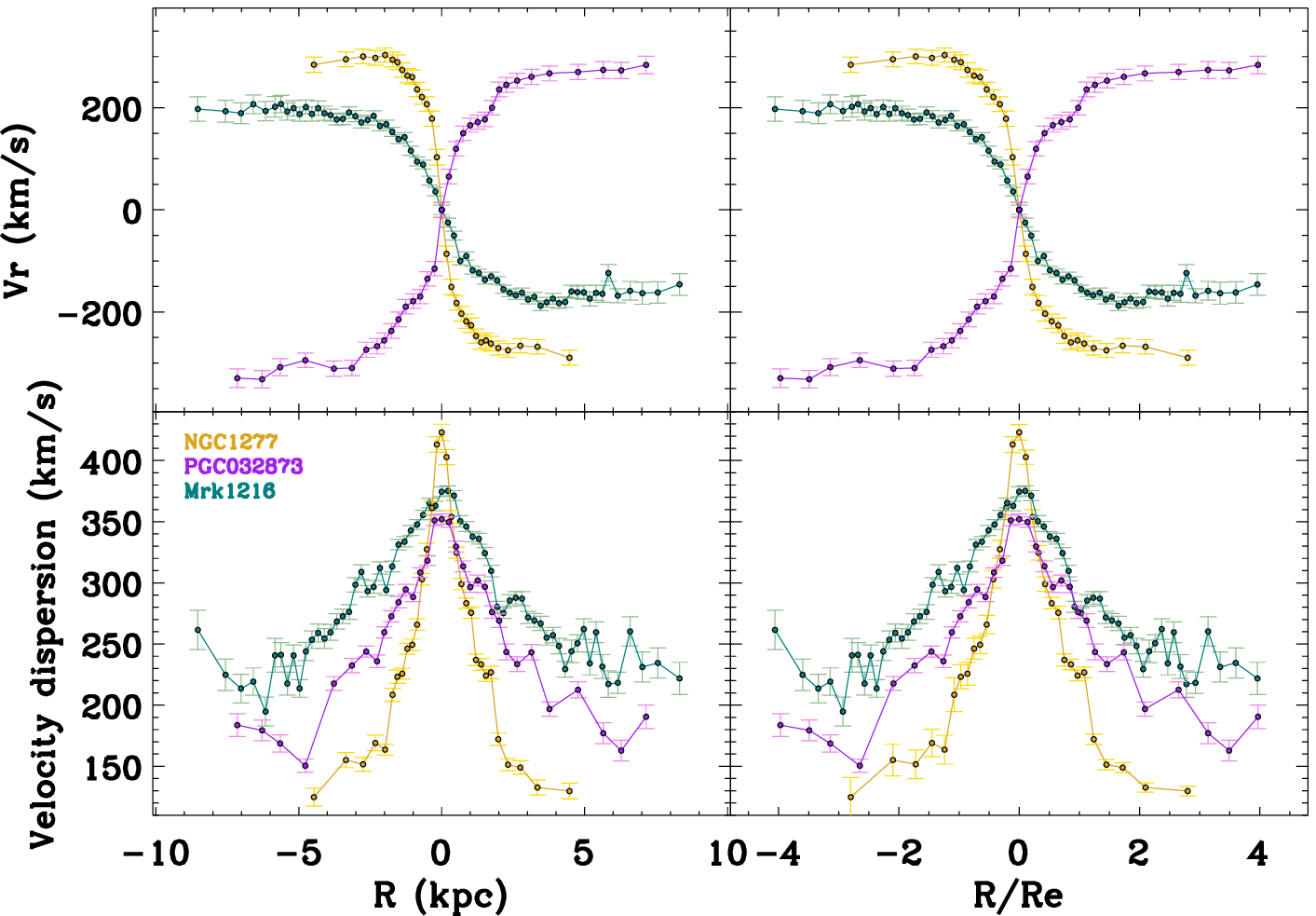}
\label{figure:2}
\caption
{Kinematic profiles obtained with {\tt pPXF}. Upper panels represent the derived rotational velocity and lower ones show the galaxy velocity dispersion. It can be seen that they are all strongly rotating systems with peaked $\sigma$ profiles, as seen for NGC\,1277 (data from T14).}
\end{figure}

\subsection{Morphology and stellar mass density profiles}
The new relic candidates PGC\,032873 and Mrk\,1216 are quite similar morphologically speaking, showing the compact elongated disky shapes seen for NGC\,1277 (Fig. 1). This is the same morphology that was predominant for the massive galaxy population at redshifts z$\sim$2 (e.g. \citealt{Buitrago2008}; \citealt{vanderWel2011}). The F160W images are significantly deeper than the F625W, showing an extended roundish component enclosing the lenticular shape on the innermost part of PGC32873 and Mrk\,1216. With these deep images we can see that there are no apparent signatures of tidal tails, asymmetries or any other features that could indicate any past or current interaction or stripping. \\
The circularized effective radii were derived using the curve of growth of each galaxy, being R$_{\mathrm{e}}$=1.8$\pm$0.2 kpc for PGC\,032873 and R$_{\mathrm{e}}$=2.3$\pm$0.2 kpc for Mrk\,1216. Figure 3 shows the location of the candidates in the stellar mass-size plane, highlighting their nature as outliers. However, note that the criteria for compactness varies from one work to another, going from the more relaxed limits of \citet{Barro2013} to the more constraining ones of T09. In order to give a meaningful sense of the compactness, we use here the criteria imposed in AFM15. That is a measure of how much the galaxies deviate from the expected size according to the stellar mass-size relation of \citet{Shen2003} as shown in Figure 3. We assume a galaxy is compact if R$_{\mathrm{e}}$/R$_{\mathrm{shen}} \lesssim$ 0.33, that corresponds to $\sim$1/3 of the normal size of a massive galaxy at z=0. These values are stated in Table 1, showing that PGC\,032873 fully complies with the compactness criteria but that Mrk\,1216 is on the limit. This will be further discussed in Section 4.\\
It was shown in T14 that the stellar mass densities of massive relics do not resemble to any of the average profiles found in local Universe galaxies, while matching almost perfectly with those of the high redshift massive population. Although our galaxies are clearly elongated, the mass density profiles shown in Figure 4 were obtained using circular apertures. This is done to allow a comparison of our effective radii with the ones that have been estimated for similar galaxies both at low and high-z. Once we obtained the F625W and F160W surface brightness profiles, we created the stellar mass density profiles of the galaxies assuming that there are no stellar population gradients throughout the entire structures. In other words, we assume that the M/L was the same through the radial structures. The applied M/L for each surface brightness profiles is the one that consistently reproduces the total stellar masses quoted in Table 1. The stellar mass density profiles of the three galaxies are all alike, being significantly denser in the inner regions compared to galaxies of similar mass but with extended sizes (i.e. the general population of present-day galaxies; see T14). Again, their stellar mass density profiles resemble the one reported for NGC\,1277, and are thus in agreement with those found for high-z massive galaxies (Fig. 4, grey lines). \\
Altogether, both their visual appearance as well as their detailed structural profiles, are almost identical to those found for massive galaxies at z$\sim$2.

\begin{figure}
\centering
\includegraphics[scale=0.42]{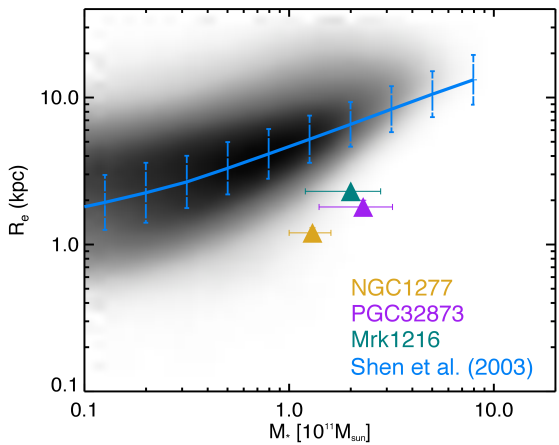}
\label{figure:3}
\caption
{Stellar mass-size distribution of the SDSS galaxies. The local mass-size relation for n$>$2.5 by Shen et al. (2003) is shown as the cyan line. The two new candidates and NGC\,1277 (filled triangles) are all clear outliers of the local stellar mass-size relation.}
\end{figure}

\begin{figure}
\centering
\includegraphics[scale=0.4]{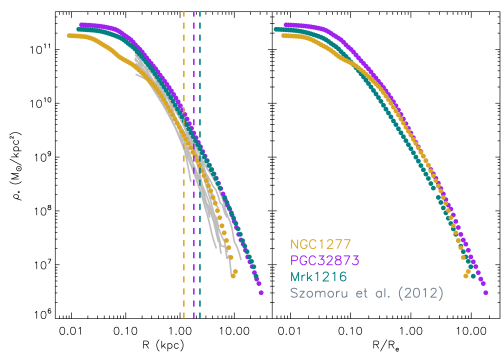}
\label{figure:4}
\caption
{Circularized stellar mass density profiles of the two relic candidates compared to the one of NGC\,1277. Overplotted in grey are the stellar mass density profiles for a sample of z$\sim$2 compact and massive galaxies from \citet{Szomoru2012}. Vertical dashed lines represent the position of the effective radii of each galaxy.}
\end{figure}

\subsection{Stellar population radial variations}
Once the criteria for being massive and compact have been both confirmed, the last criteria that a massive relic candidate has to meet is to have stellar populations that are old  ($\gtrsim$10\,Gyr) along its entire structure. We next study their Star Formation Histories (SFHs) and stellar populations (ages and metallicities) up to several galactocentric distances. We employ the extended version of the MILES stellar population synthesis (SSP) models, E-MILES (\citealt{Vazdekis2015}; \citealt{Vazdekis2016}), which cover a wide range of ages (0.03$\le$t$\le$\,14.0\,Gyr), metallicities (-2.27$\le$[M/H]$\le$\,+0.27) and IMF slopes (0.3$\le \Gamma_\mathrm{b} \le$3.3). We derive the relevant properties using both the full-spectral fitting technique with the code {\tt STARLIGHT} (\citealt{CidFernandes2005}, \citealt{CidFernandes2010}) and the classical approach of line-strength measurements. \\

\subsubsection{Initial mass function}
There is, however, a parameter that we must have in mind before performing any stellar population analysis. There has been recently a great deal of works pointing out that the IMF might not be a universal parameter. The main property responsible for this has been claimed to be the galaxy velocity dispersion, as a proxy for the galaxy mass (e.g. \citealt{Cappellari2012}, \citealt{Ferreras2013}; \citealt{LaBarbera2013}). Under this hypothesis, galaxies with steep velocity dispersion profiles, like massive galaxies, should also show a radial variation in their IMF slopes. Indeed, this has been hinted for a handful of individual ETGs (e.g. \citealt{MartinNavarro2015a}; \citealt{LaBarbera2016}). However, this is not the case for the massive relic NGC\,1277 \citep{MartinNavarro2015c}, for which a negligible radial IMF variation was found. While the results were compatible with a very steep IMF for the whole structure of NGC\,1277, the lack of IMF gradients pointed out to another property being the main driver for such variations (e.g. $\alpha$-enhancement or metallicity; see \citealt{vanDokkum2012} or \citealt{MartinNavarro2015c}). \\
In order to deepen our understanding of the impact of the IMF in massive relic galaxies, we use new deep spectroscopic data obtained with FOCAS at Subaru Telescope. We derive the IMF slope that better fits the two new relic candidates following the approach described in \citet{LaBarbera2013} and \citet{MartinNavarro2015b}, while studying how it varies along their structures. First, the equivalent width of age (H$\beta{o}$), metallicity ([MgFe]') and IMF (e.g. bTiO, TiO$_1$, CaT) sensitive features are measured using the standard line-strength technique. After correcting from possible nebular emission in the Balmer lines, we minimize a $\chi^2$ merit function with age, metallicity and IMF slope as free parameters (see Eq.\,1 in \citealt{MartinNavarro2015b}). The abundance pattern correction is done in two ways, either using the local velocity dispersion (see La Barbera et al. 2013 for details) or treating it as an additional free parameter. Given the strong degeneracies affecting the IMF measurement, we repeated the fitting process using different sets of indices. The final IMF gradients presented here are the average values of all the different fits (e.g., with the two abundance pattern corrections plus the best-fitting solutions based on different index combinations). \\
Unfortunately, the slight difference in redshift between Mrk\,1216 and PGC\,032873 prevented us from using a homogeneous set of line-strength indices. Important indicators used in previous works, such as the TiO$_2$ molecular band or the Na\,8190 index, were not usable in this case, as the telluric correction (both in absorption and emission) was not accurate enough at the level required by a robust IMF analysis. For Mrk\,1216 we based our analysis on the bTiO, TiO$_1$ and CaT spectral features. The more distant PGC\,032873 was more problematic and only the Ca1 line of the CaT tripled could be used. Note that despite these differences, here we are focusing in the radial behavior of the IMF, which is compatible for all three galaxies.\\
The derived IMF gradients are shown in Figure 5, being consistent with previous observations of NGC\,1277 \citep{MartinNavarro2015b}. It is clear that the central regions of our massive relic candidates host an enhanced fraction of dwarf stars (i.e. a steeper than universal Kroupa IMF). When moving towards the outskirts, this dwarf-to-giant ratio slightly decreases. Although these new data reaches further than previously for NGC\,1277, we find that the inferred IMFs are steep, reaching a plateau after $\sim$2\,R$_{e}$ with $\Gamma_\mathrm{b}>$ 2.0. \\

\begin{figure}
\centering
\includegraphics[scale=0.95]{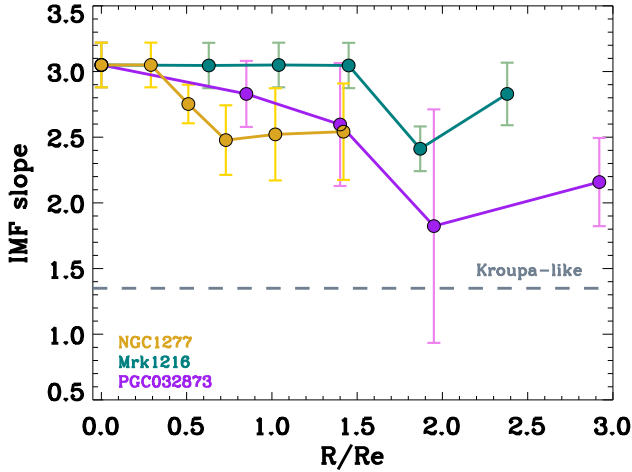}\\
\label{figure:5}
\caption
{Radial IMF-slope gradients for the two relic candidates (this work) and for NGC\,1277 as in \citet{MartinNavarro2015b}. The vertical axis indicates the slope of the IMF in the bimodal case ($\Gamma_\mathrm{b}$). For reference, the Milky-Way approach is shown as a horizontal grey dashed line. Both PG032873 and Mrk\,1216 show mild IMF radial gradients, being very bottom-heavy out to $\sim$1.5R$_{\mathrm{e}}$, similarly to NGC\,1277. The larger radial coverage of the new data allows us to reach almost up to 3R$_{\mathrm{e}}$ for these two new candidates, showing that although the inferred IMF decreases with galactocentric distance, they all seem to reach a plateau with $\Gamma_\mathrm{b}>$ 2.0.}
\end{figure}

\subsubsection{Star Formation Histories}
We showed in \citet{FerreMateu2013} the strong impact that using a varying IMF can have in the derived stellar population analysis, in particular when looking at the SFHs of galaxies. To account for these variations and thus obtain the most accurate stellar population estimates, we use the IMF slope inferred in the previous subsection for all three galaxies (the previous analysis of T14 was performed assuming a standard Kroupa IMF). We also include a universal Kroupa IMF analysis, for comparison purposes with previous works.\\
We investigate how the galaxies built-up their stellar mass through cosmic time in order to determine the timescales of galaxy formation. For this means, we first derive the SFHs within 1R$_{\mathrm{e}}$ for the three galaxies. They all hold high resemblance, showing virtually old ages and single-burst like SFHs. This is shown in Figure 6, which represents the cumulative mass fraction that is built up for each galaxy. Using a steeper IMF slope in the case of NGC\,1277 (left panel, yellow line) does not render different results to those in T14 (grey line), reinforcing the nature of NGC\,1277 as the first, fully confirmed massive relic galaxy. It was shown that such assembly history implied very short formation timescales, of the order of hundred Myrs. This is compatible with the new star formation histories derived here. Note that we are unable to resolve age differences of less than the first 0.5\,Gyr. In that sense, we find that NGC\,1277 built up more than 50\% of its stellar mass within the first 0.5\,Gyr and reached 90\% before 1\,Gyr had elapsed. PGC\,032873 (middle panel) replicates this extremely short formation timescale. In this case, using a standard or steeper IMF has almost no impact in the derived SFHs, which shows always a single-burst like shape. Mrk\,1216 is the only one of the candidates to deviate from this extreme formation histories, even when using the standard IMF assumption. Although it also formed the bulk of its stars at the earliest epochs in a fast event, it shows a slight delay and a more extended accretion history, assembling its mass 1\,Gyr later than NGC\,1277 and PGC\,032873. \\
These results are compatible with the timescales inferred from the measurement of the $\alpha$-enhancement derived from the absorption line indices (e.g.  \citealt{deLaRosa2011}; \citealt{McDermid2015}). Unfortunately, similarly to what happened in T14, extreme extrapolations in the index-index model grids are necessary to derive these quantities, and  the inferred values are always far beyond the limit of the SSP grids. Therefore, we can only consider them as a lower limits of +0.40\,dex. But even with these lower estimate, a very short timescale of a fem Myrs is inferred. \\
All three objects show early peaked, fast star formation events within their 1R$_{\mathrm{e}}$.  Under the two-phase formation scenario considered here, this would represent the first monolithic-like phase that created the massive, compact core, which would grow in size by the accretion of smaller satellites at later times being deposited in the outskirts ($\gtrsim$ 1R$_{\mathrm{e}}$). In order to be truly confirmed relics, we must ensure there are no signs of such events at later epochs. We do so by studying the stellar content with increasing annular bins, as shown in the Appendix. We confirm that NGC\,1277 stands firmly as a massive relic galaxy even using steeper IMF profiles and both PGC\,032873 and Mrk\,1216 show SFHs compatible with the relic galaxy assumption. Minor differences appear as we move to larger galactocentric distances. To better understand them, Figure 7 shows the radial profiles of the stellar population properties. Upper rows show the mean mass-weighted ages, followed by the fraction of stars (in mass) with old stellar populations ($\gtrsim$10\,Gyr), and the mean mass-weighted metallicity on the third row. They show that mean ages and SFHs remain old and with virtually no variations. Note that according to \citet{FerreMateu2013}, a steep IMF slope allows for more extended SFHs than if a universal IMF is used (grey lines in Figure A1). Since the most extreme IMF slopes are restricted to the core of our galaxies, where the SFHs are the most monolithic-like, a varying IMF will mildly reshape the inferred SFHs but will keep them entirely old (as shown in Figure A1). In the outskirts, where the fraction of relatively younger (but always older than 8 Gyr) stellar populations starts to show, the IMF slope is no longer as exotic, thus the differences are expected to be smaller. This explains why the mean age of NGC\,1277 seems to increase, but it also shows that even in the outermost distances (and thus the lowest IMF slopes) this galaxy had a genuinely single-burst like episode of formation. Therefore, our finding of a mild IMF gradient does not significantly affect the relic character of our sample.\\
The only noticeable variation that can be seen is a mild gradient in the mass-weighted mean metallicities. The innermost parts of the galaxies are saturated ([Z/H]$>$0.26\,dex) and thus must be considered as a lower limit. In  any case, both candidates have very high metallicities up to at least 1R$_{\mathrm{e}}$, which then slightly decline with galactocentric distances. Therefore, in general trends, the stellar populations of the massive relic candidates do not differ particularly to those obtained for massive ETGs \citep{SanchezBlazquez2007}, although it is not possible to determine the real gradient due to the saturation in the central parts.

\begin{table}
\label{table:2}                      
\centering
\caption{Stellar populations properties}    
\begin{tabular}{c | c c  c c}   
\hline\hline      
                 &    IMF slope   & Age &   Fract. old & Metallicity  \\  
                 &                        &(Gyr) &      $\%$                        &  (dex)          \\  
\hline  
NGC\,1277      & 2.5 & 13.7$\pm$1.6 & 100  & 0.264$\pm$0.055\\
PGC\,032873  & 2.8 & 12.7$\pm$1.6 &  99   & 0.262$\pm$0.048\\
Mrk\,1216        & 3.0 & 12.8$\pm$1.5 &  99   & 0.259$\pm$0.052\\
 \hline                                
\end{tabular}

{Summary of the derived stellar population properties within 1R$_{\mathrm{e}}$. Column 2 shows the IMF slope considered for each candidate. Columns 3, 4 and 5 show, respectively, the mean mass-weighted age (in Gyr), the fraction of stellar populations created with $\gtrsim$10\,Gyr and the mean mass-weighted metallicity, all derived with the full-spectral-fitting technique.}
\end{table}

\begin{figure*}
\centering
\includegraphics[scale=0.95]{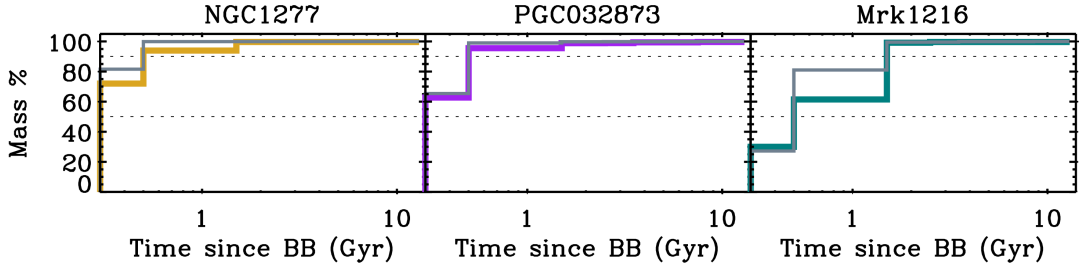}\\
\label{figure:6}
\caption{Cumulative mass fraction within 1R$_{\mathrm{e}}$ for our two candidates and NGC\,1277. They have been inferred from the star formation histories accounting for a non-universal IMF slope of 3.0 for Mrk\,1216, 2.8 for PGC\,032873 and 2.5 for NGC\,1277 (colored lines), although the cumulative fraction assuming a universal IMF approach is also shown in grey solid line. The time axis corresponds to the time since the Big Bang, being t=0 the equivalent to the oldest SSP in the models (14\.Gyr). It is clear that all three galaxies formed their stars very early on in very short timescales, although a longer episode is seen for Mrk\,1216, reaching 90\% of its mass almost 1\,Gyr later than PGC\,032873 and NGC\,1277.}
\end{figure*}

\begin{figure}
\centering
\includegraphics[scale=0.59]{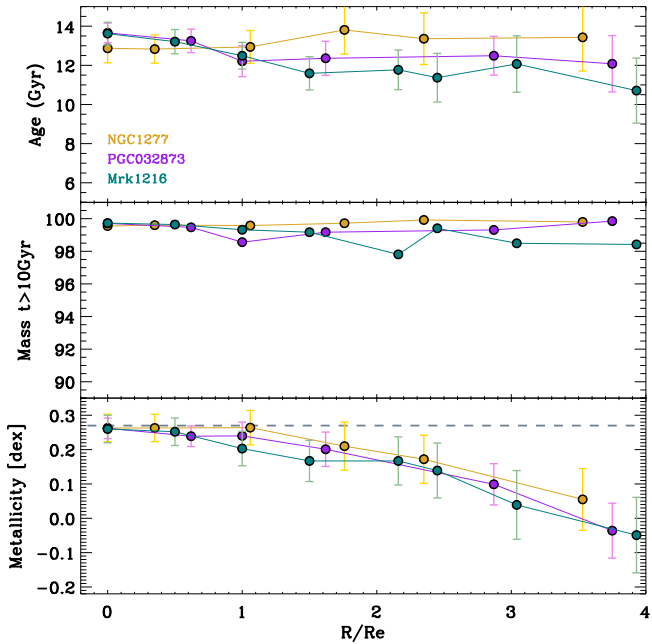}\\
\label{figure:7}
\caption{Radial profiles of the stellar population properties derived from the full-spectral-fitting approach for each galaxy. For each aperture, the correspondent IMF slope from Fig.\,5 has been considered. While both the mass-weighted age (top) and fraction of stars older than 10\,Gyr (middle panel) remain almost unchanged, the total metallicity (bottom) is the only parameter that strongly varies, decreasing as we go to further galactocentric distances, as it was previously found for NGC\,1277.}
\end{figure}

\section{Discussion}
It is straightforward to see that all the results obtained for PGC\,032873 and Mrk\,1216 closely follow the trends seen previously for NGC\,1277, which in turn, mimic those seen for high redshift massive galaxies. In fact, the central parts of the three galaxies present almost identical properties: their SFHs show early and peaked formation events, with very short timescales and old mean mass-weighted ages ($\sim$13\,Gyr). It is only after 1R$_{\mathrm{e}}$ that small differences start to show up, mostly in their metallicities. Mrk\,1216 also shows a noticeably more extended SFH that accounts for a delay on its formation timescale. But altogether, they are set to pertain to the much sought massive relic family. \\
It is important to stress that what seems to make these relic galaxies unique, rather than their stellar populations or stellar masses, are their extreme morphologies and dynamics. They have strongly peaked high velocity dispersion profiles with high radial velocities, compact and elongated morphologies with no clear signs of interactions and show density profiles that are highly concentrated in the innermost parts. This set them apart of the typical z$\sim$0 massive ETGs and instead, they represent the properties from the in-situ phase.\\
Considering that the furthest of these galaxies lies at 106\,Mpc, the volume enclosed by a sphere with such radius would be of $\sim$5$\times$10$^{6}$Mpc$^{3}$. With 3 massive relic galaxies confirmed to date, this would imply a number density of 6$\times$10$^{-7}$Mpc$^{-3}$. This represents a lower limit, as there might be other relic galaxies within that volume that have not yet been discovered. It is worth noting that this estimate falls within the numbers predicted by the $\Lambda$CDM model \citet{Quilis2013}.

\subsection{Massive relics and their SMBHs}
There is yet another peculiar characteristic that massive relic galaxies seem to have: they seem to host extremely large SMBHs in their centers (e.g. \citealt{vandenBosch2012}; \citealt{vandenBosch2015}; \citealt{Walsh2016}; \citealt{Yildirim2015}). These have been also nick-named \"ubermassive black holes and studied in AFM15. In the latter, we proposed an scenario to explain the nature of such monsters, showing that there is nothing wrong with their SMBHs, but they are a natural consequence of the unusual path massive relic galaxies undergo. We showed that, if massive relic galaxies are allowed to grow in mass, velocity dispersion and size the amount expected during the accretion phase via mergers, i.e. they follow the expected path for normal massive ETGs (e.g. \citealt{Oogi2013}; \citealt{Wellons2015}), these variations would place them closer to the present-day scaling relations. The fact that the property mostly affected by such unusual path is the galaxy stellar mass further reinforces this possibility. We know from semi-analytical models and cosmological simulations that an individual massive galaxy can increase by almost $\sim$7 times its size during the merger phase after z=2, while its velocity dispersion is not expected to vary (at most a factor $\sim$1.1 ; \citealt{Hilz2012}). This is consistent with our candidates not showing strong deviations from the local scaling relations in the M$_{\bullet}$-$\sigma$ plane. However, the amount of stellar mass that such galaxies can incorporate during that phase can account for up to a $\sim5$ times increase (e.g. \citealt{Oser2010};  \citealt{Trujillo2011}).\\
Figure 8 shows the upper mass end of the SMBH-host galaxy mass scaling relation. It includes all the published SMBHs mass estimates considered in AFM15 (grey points; \citealt{McConnell2013}; \citealt{Kormendy2013}; \citealt{Graham2013}), the sample of relic candidates from AFM15 (black dots), the two candidates studied in this work, and NGC\,1277 (filled triangles). The fit for all galaxies from \citet{McConnell2013} is shown in a solid line, with its 1, 3 and 5$\sigma$ deviations. Both PGC\,032873 and NGC\,1277 were used previously in AFM15, as they both had available SDSS spectra. We now include Mrk\,1216, using the upper limit log(M$_{\bullet}$/M$_\odot$)=10 obtained in \citet{Yildirim2015} from a detailed dynamical modeling. We also use their latest measurement for NGC\,1277, which is now reported to be log(M$_{\bullet}$/M$_\odot$)=10.1. Although this is considerably lower than the previous log(M$_{\bullet}$/M$_\odot$)=10.23 \citep{vandenBosch2012}, this galaxy remains as one of the most extreme outliers, almost at a 5$\sigma$ level. The two new relics show more than a 3$\sigma$ deviation, with Mrk\,1216 being the less extreme of all. In fact, if we adopt the latest measurements of Mrk\,1216 reported by \citet{Walsh2016} (small green triangle) this galaxy is even less extreme, while still being an outlier. This is compatible with the proposed scenario of AFM15, where a relation between how much of an outlier the relics are and the extent of the star formation histories of the galaxies. While PGC\,032873 does show a formation history resembling to a single-burst episode, as we go to the outer parts this episode is seemingly longer. This hints to a later suppression of star formation compared to NGC\,1277 (as seen by Fig.\,6), having time to accumulate a bit more of stellar mass and therefore being less of an outlier. On the contrary, Mrk\,1216, the less of our outliers, turns out to be the one with a more extended SFH.\\
We should mention that there has been some recent works pointing out that these objects would not be as much outliers if another M$_{\bullet}$-M$_{*}$ relation is considered (e.g. \citealt{Graham2016a}; \citealt{Graham2016b}; \citealt{Savorgnan2016a}; \citealt{Savorgnan2016b}). By including spiral galaxies, \citet{Savorgnan2016a} showed that two separate fits should be considered, one for the low-mass spheroidals and the other for the massive ones. The later, which would correspond to our object type, is shown with a blue line in Fig. 8. We see that the effect at the higher mass regime is basically a shift in the correlation, placing our outliers closer to the scaling relation. There are other facts that could change the position of our objects in the local scaling relations. Better black hole measurements for all these objects are crucial, as it has been proven for NGC\,1277 and recently Mrk\,1216 (\citealt{Yildirim2015}; \citealt{Walsh2016}). In addition, assuming the steeper IMFs derived in Section 3.3.1 would increase out stellar masses estimates by a factor of $\sim$1.6 \citep{FerreMateu2013}. Also, such IMF variation should be accounted to compute the SMBH mass (see e.g. \citealt{Laesker2013}). A study of the IMF impact on such scaling relations is out of the scope of this paper, but it is important to highlight that none of these effects would have relevant implications for the scenario being tested here nor for the nature of our objects. If anything, if these objects were less outliers it would alleviate the amount of evolution required during the second phase, which is quiet extreme within the current \citet{McConnell2013} scaling relations. \\
The scenario presented in AFM15 and further confirmed here implies that all relic galaxies should have black holes that are larger than expected by the local scaling relations, regardless of galaxy mass. How extreme or unusual the black holes are, is still to be confirmed by taking into account the stated caveats.

\begin{figure}
\centering
\includegraphics[scale=0.65]{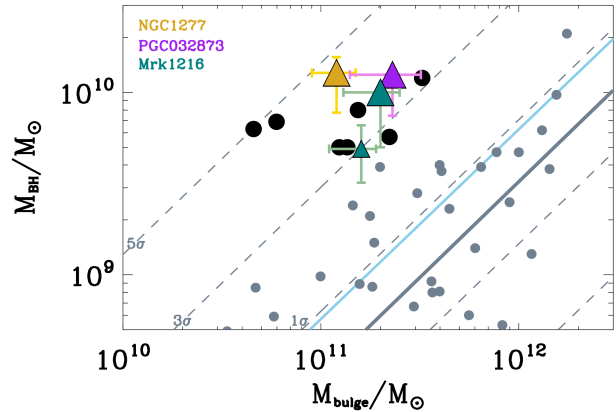}\\
\label{figure:8}
\caption{Black hole mass scaling relation with bulge mass. Black dots correspond to the \"UMBH objects from AFM15 (excluding NGC\,1277 and PGC\,032873), while grey dots correspond to the compilation of published galaxies with SMBH detections. The three confirmed relic galaxies are shown in colored triangles. The black solid line is the fit corresponding to McConnell $\&$ Ma (2013) and the dashed lines limit the 1$\sigma$, 3$\sigma$ and 5$\sigma$ deviations from it. The blue solid line shows the newly defined scaling relation from Savorgnan et al. (2016a), which would place the relic galaxies already closer to the relation and thus, invoking less mass evolution would be needed. Another way to alleviate the extreme outliers is by having more detailed dynamical modeling of the SMBHs, which will render more accurate mass measurements, as shown by the small green triangle, the latest SMBH mass estimate for Mrk\,1216 \citep{Walsh2016}.}
\end{figure}

\subsection{Environmental impact on massive relic galaxies}
While studying the different properties for each galaxy there seems to be a consistent pattern. NGC\,1277 is usually the most extreme case: it is the most compact, it shows the less variation in its SFHs, has the shortest formation timescales and thus seems to be the most relic-like of the three. At the same time, it is the most extreme outlier in the SMBHs scaling relations. Mrk\,1216, instead, seems to be at the other end: it is the less outlier in all the scaling relations, it presents a larger size and it built up its mass in a slightly larger timescale (by $\sim$1\,Gyr difference), showing a later accretion history. PG032873 seems to be always in between, sometimes resembling more to NGC\,1277, sometimes more to Mrk\,1216. \citet{Yildirim2015} pointed out that the dynamical properties they derive for Mrk\,1216, if compared to those of NGC\,1277, seem to indicate that Mrk\,1216 is a good candidate for a massive relic that "has already entered a path of becoming a regular, fast-rotating elliptical". Our findings support this claim, suggesting a correlation between how much of a relic a galaxy is compared to how extreme its properties are. In other words, we are catching galaxies at different stages of the relic path.\\
But, is there really such a \textit{"degree of relic"}? If the three galaxies studied here are, in general, so similar and clearly pertaining to the same family, what makes them to be at different points on the massive relic track?  Local environment might be the clue. NGC\,1277 resides close to the center of Perseus, the most massive nearby cluster. It was not surprising to find this first relic galaxy in such rich environment, as galaxy clusters are probabilistically the easiest place to find the elusive massive relics (e.g. \citealt{Poggianti2013b}; \citealt{Shankar2014}; \citealt{Stringer2015}; \citealt{Damjanov2015b}, \citealt{Volonteri2016}). However, relic galaxies are also expected to be found in less dense environments (e.g. \citealt{Poggianti2013}; \citealt{Damjanov2014};  \citealt{Ma2014}; \citealt{Peralta2016}), as our newly confirmed objects. According to the hierarchical $\Lambda$CDM paradigm, dense and dynamical environments would be responsible for starting the galaxy formation earlier and faster. This would give a heads-up start to the galaxies inhabiting those regions, which would show the older ages. At the same time, the active intracluster environment would inhibit any posterior star formation events, either via stripping or heating of cold gas. This would render the most extreme cases of relics, such as NGC\,1277. On the contrary, a less rich environment would start the star formation slightly later \citep{Thomas2005} and it would not be strong enough to prevent posterior (even if rare) encounters or star formation events. This would make field massive galaxies to show delayed and/or more extended star formation histories, indicative of some minor encounters with other galaxies. This could also explain why, on average, galaxy sizes in clusters are smaller than in the field (at a fixed stellar mass; e.g. \citealt{Cebrian2014}). \\
Following this idea, NGC\,1277 would then be the perfect and most extreme case of a massive, pristine relic, while Mrk\,1216 would be an example of a galaxy that pertained to that class but that started it journey to become a larger, regular ETG. If we follow the most strict criteria to select massive relic candidates (like in \citealt{Trujillo2009}), then Mrk\,1216 should not be included, as both the SFH and the size are slightly off the limits. However, with the idea of a \textit{"degree of relic"}, where these peculiar objects would be allowed to have a range of properties, Mrk\,1216 would comply. This also works in better agreement with other studies where their criteria for selecting candidates are more relaxed (i.e. allowing up to $<$10$\%$ of mass evolution, as in \citealt{Quilis2013} or \citealt{Poggianti2013}).\\
The environment seems to determine how much of a relic a galaxy is, or how much into the path of becoming a more normal massive ETGs it has gone. In fact, this environment dependence could also explain the nature of the sample of massive and compact local galaxies from T09 that were first considered as massive relic candidates. Sharing the extreme morphological and dynamical signatures of the high-z red nuggets, these galaxies were first excluded due to varying amounts of recent star formation \citep{FerreMateu2012}. Interestingly, they all live in low density environments. Therefore, under the \textit{"relic degree"} assumption presented here, those varying amounts of star formation could represent later stages on how far these galaxies have walked away from the pristine relic properties, shedding some light into their mysterious origin.

\section{Summary}
We have confirmed two new massive relic galaxies by studying in detail their morphologies, density profiles, stellar kinematics and stellar content at different galactocentric distances. \\
PGC\,032873 and Mrk\,1216 both show the disky-like compact sizes and centrally-dense stellar mass profiles seen for the z$\sim$2 red nuggets, with highly symmetrical shapes with no signs of tidal streams that could indicate that part of the stellar mass had been stripped. In terms of their kinematics, both objects closely resemble the profiles seen for NGC\,1277, the first confirmed massive relic galaxy (T14). They have strong radial velocities ($>$200$\mathrm{km\,s^{-1}}$) and steep velocity dispersion profiles peaking at $\sim$350$\mathrm{km\,s^{-1}}$. \\
They both have large stellar masses (M$_{*}\sim$2$\times$10$^{11}$ M$_\odot$), but those are not enough to account for the even larger SMBHs they host in their centers. This has allowed us to further prove that the massive relic galaxy family should be found as outliers in the local scaling relations due to their unusual path of formation (i.e. skipping the accretion phase).\\
A detailed analysis of their stellar content and how each galaxy built it up over time has given us the last confirmation about their relic nature. We have considered the impact of using a non-universal IMF slope for our stellar population analysis by first analyzing the radial IMF variations of the candidates. They both show steep IMF profiles along their entire structure, going from a slope of $\Gamma_\mathrm{b}$=3.0 at their centers to one of $\sim$2.5 for the outermost radii. Considering these IMF estimates, we find that both PGC\,032873 and Mrk\,1216 show old ages along their entire structure ($\sim$13\,Gyr up to $\sim$4R$_{\mathrm{e}}$) and confirm that NGC\,1277 stands as a relic galaxy even using these steeper slopes. They are all compatible with having formed the bulk of their stellar content at the earliest epochs, in a very fast, single-burst like event that would represent the first phase of the depicted formations scenario (i.e. in-situ, monolithic-like phase). There are no traces of posterior events in their outer regions, showing that the confirmation as relic galaxies is robust. In terms of their metallicity, they both show radial gradients compatible with normal ETGs, although we can not determine the real steepness of the profiles due to the saturation in the central parts. Their abundance of $\alpha$-elements are compatible with the extremely short formation timescales, although we can not give a quantitative estimate due to the large extrapolations the method suffers.\\
\\
Finally, we find that the nature of the elusive and scarce family of massive relic galaxies seems to be more a morphological and dynamical issue, while their stellar population properties can only tell us how far they are in the path to become a normal massive ETG. It seems that the \textit{degree of relic} is related to the local environment where galaxies reside, with those in clusters showing the most extreme, pristine properties (smallest sizes, shortest SFHs, largest SMBHs, etc) and those in less rich environments showing less extreme properties and slightly evolved.\\
Unfortunately, this work has been done with only three fully characterized candidates. It would be now interesting to go from single anecdotic examples to larger samples to robustly test the presented hypothesis, for which large statistical and spatially resolved samples of massive relic candidates will be needed. 

\section*{Acknowledgements}
We thank the referee for their useful comments that helped improve the article greatly. This work was mostly supported by the Japan Society for the Promotion of Science (JSPS) Grant-in-Aid for Scientific Research (KAKENHI) Number 23224005 and also by the Spanish Ministerio de Econom\'ia y Competitividad (MINECO; grants AYA2009-11137, AYA2011-25527 and AYA2013-48226-C3-1-P). This article is based on observations made with the WHT operated on the island of La Palma by the Isaac Newton Group of Telescopes in the Spanish Observatorio del Roque de los Muchacho. Also observations from Subaru Telescope at the summit of Mauna Kea are presented here. The authors wish to recognize and acknowledge the very significant cultural role and reverence that the summit of Mauna Kea has always had within the indigenous Hawaiian community.  We are most fortunate to have the opportunity to conduct observations from this mountain.


\bibliography{biblio_relicsall}
\bibliographystyle{mn2e}

\appendix

\section{Radial Star Formation Histories}
We present the radial SFHs for Mrk\,1216, PGC032987 and NGC\,1277 for each annular aperture, chosen to maximize the S/N obtained using the code {\tt SNratio} from {\tt REDUCEME}. Each aperture has been treated with the derived IMF slope shown in Fig.\,5, although the universal assumption is also shown. As we only reach $\sim$2R$_{\mathrm{e}}$ for the IMF analysis, we assume the lowest IMF derived for the rest of apertures. Except for the outermost radii, which are on the limit of the optimal S/N optimal for the full-spectral-fitting technique ($\sim$20), all the apertures show, on average, that our candidates present old stellar populations, consistent with being formed 10\,Gyr ago. A slightly more extended SFH is derived for Mrk\,1216, pointing to some degree of recent activity/interaction, in particular at larger galactocentric distances. Again, all three galaxies show almost identical properties, in particular within 1R$_{\mathrm{e}}$, and all three galaxies can be considered relics of the early universe massive population. As discussed in Section 4, the small differences in the formation timescales can be related to a degree of relic that these galaxies seem to have. 

\begin{figure*}
\centering
\hspace{0.8cm}\includegraphics[scale=0.45]{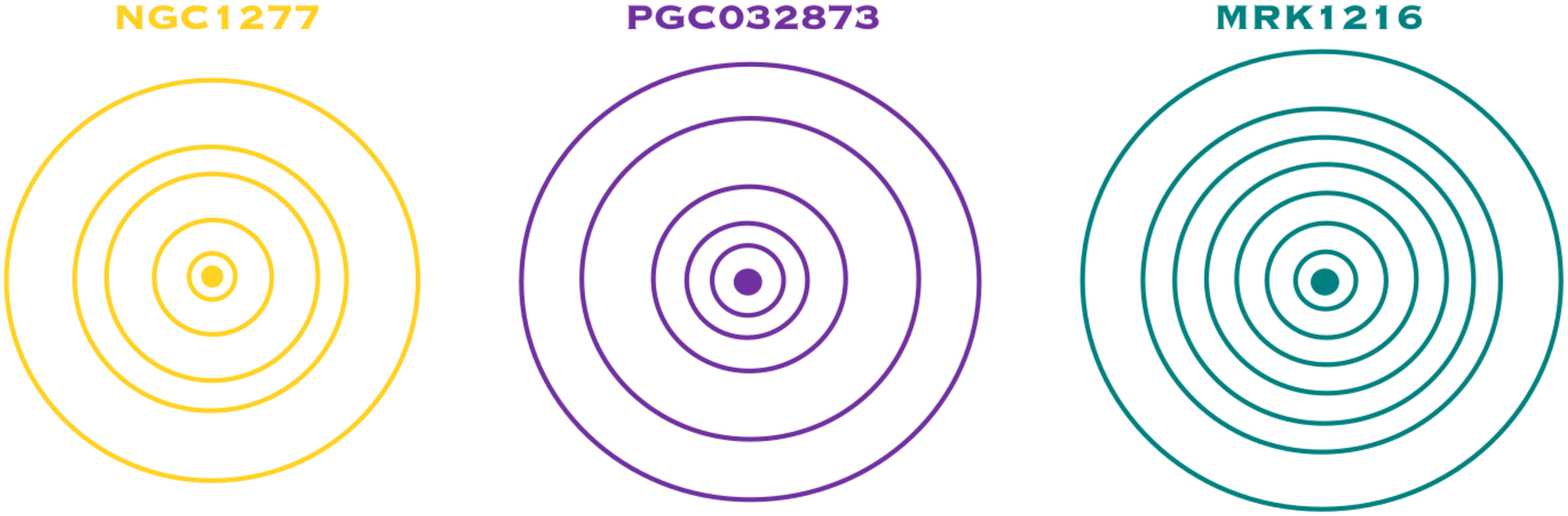}
\includegraphics[scale=0.95]{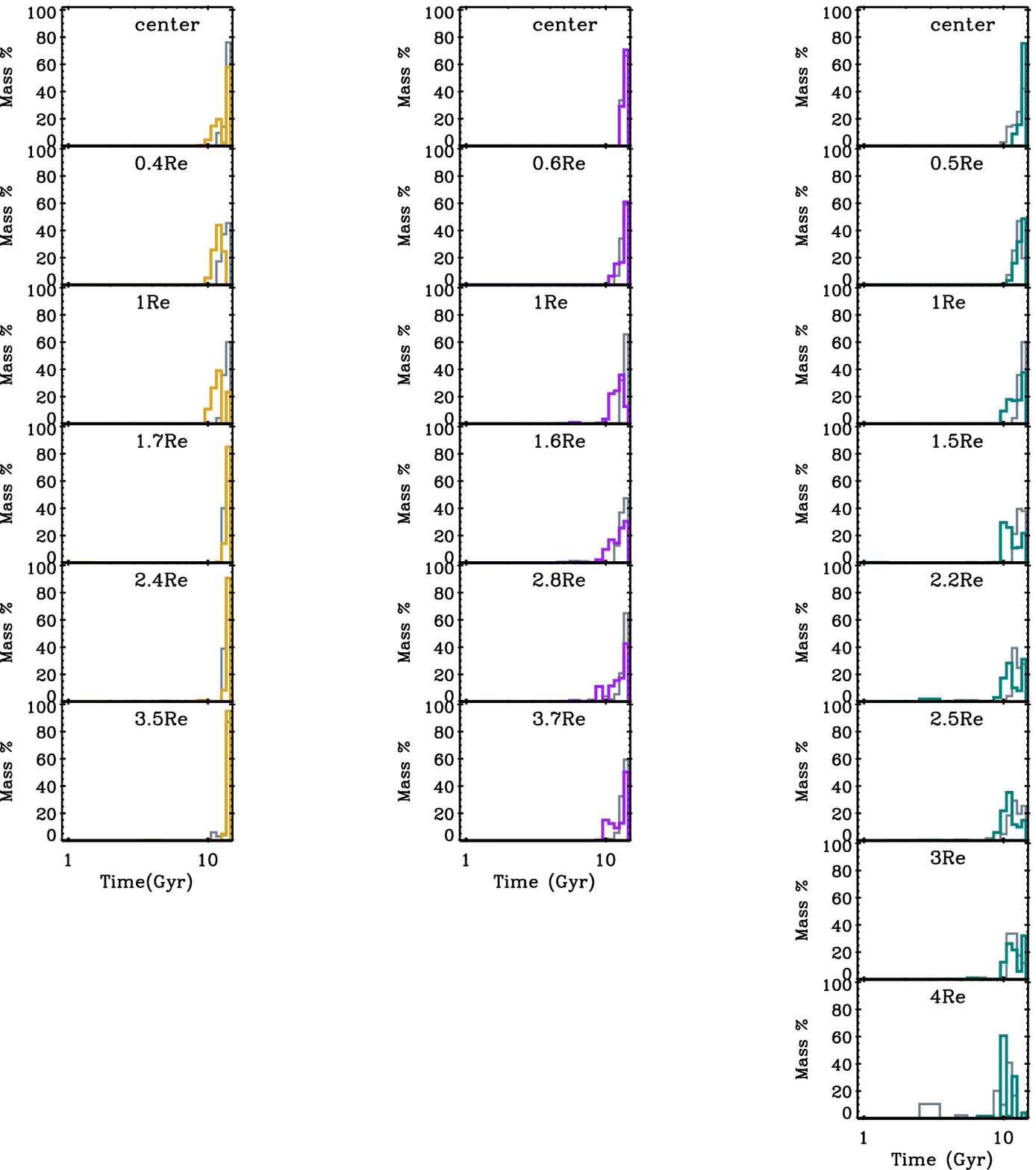}

\label{figure:a1}
\caption{Derived Star Formation Histories for each annular aperture, as shown in the upper scheme, with the apertures being shown to scale with the effective radii. Each aperture SFH is measured with the IMF slope according to Fig.\,5, but the Kroupa approach is also plotted in grey histograms. }
\end{figure*}


\bsp	
\label{lastpage}
\end{document}